\documentclass[showpacs,showkeys,twocolumn,preprintnumbers,superscriptaddress,amsmath,amssymb,nofootinbib]{revtex4}

\usepackage[dvips]{graphicx}
\usepackage{epsfig}
\usepackage{color}
\usepackage{latexsym}
\usepackage{amsmath}
\usepackage{amssymb}
\usepackage{eufrak}
\usepackage{euscript}
\usepackage{pstricks}
\usepackage{picture}
\def\[{\left\lbrack}
\def\]{\right\rbrack}

\def\({\left(}
\def\){\right)}

\newcommand{\bee}{\begin{equation}}
\newcommand{\eee}{\end{equation}}
\newcommand{\eaa}{\end{eqnarray}}
\newcommand{\baa}{\begin{eqnarray}}

\def\ni{\noindent}

\def\no{\nonumber}

\begin{document}

\title{\Large Emergence of cosmic space, Gauss-Bonnet gravity, \\ MOND theory and nonextensive considerations}

\author{Everton M. C. Abreu}\email{evertonabreu@ufrrj.br}
\affiliation{Grupo de F\' isica Te\'orica e Matem\'atica F\' isica, Departamento de F\'{i}sica, Universidade Federal Rural do Rio de Janeiro, 23890-971, Serop\'edica - RJ, Brasil}
\affiliation{Departamento de F\'{i}sica, Universidade Federal de Juiz de Fora, 36036-330, Juiz de Fora - MG, Brasil}
\author{Jorge Ananias Neto}\email{jorge@fisica.ufjf.br}
\author{Albert C. R. Mendes}\email{albert@fisica.ufjf.br}
\author{Daniel O. Souza}\email{danieljf@fisica.ufjf.br}
\affiliation{Departamento de F\'{i}sica, Universidade Federal de Juiz de Fora, 36036-330, Juiz de Fora - MG, Brasil}

\date{\today}

\begin{abstract}
\noindent In this paper, by using Verlinde's formalism and a modified Padmanabhan's prescription, we have obtained the lowest order quantum correction to the gravitational acceleration and MOND-type theory by considering a nonzero difference between the number of bits of the holographic screen and the number of bits of the holographic screen that satisfy the equipartition theorem (the bulk). We will also carry out an analysis for the pure and an asymptotic (actual) de Sitter Universe considering the holographic principle.  We had also used nonextensive concepts into the theory and we accomplished a $N$-dimensional generalization of our results.  Some physical consequences of the nonextensive ideas in Gauss-Bonnet (GB) gravity theory were analyzed also.  We have obtained the $q$-parameter as a function of the GB coefficient and some physical aspects were discussed. 
\end{abstract}


\maketitle

\section{Introduction}

The formalism proposed by E. Verlinde \cite{Verlinde} obtains the gravitational acceleration  by using the holographic principle and the equipartition law of energy. His ideas relied on the fact that gravitation can be considered universal and independent of the details of spacetime microstructure.  Besides, he brought new concepts concerning holography since the holographic principle must unify matter, gravity and quantum mechanics. It is important to mention that similar ideas have also been given by Padmanabhan \cite{Pad2}. 

An important concept used by Verlinde is the notion of bits which can be  understood as the smallest unit of information in a holographic screen. Indeed, bits play an essential role in Verlinde's formalism because when the total bits number is assumed to satisfy completely the equipartition law of energy, we have the well known formula of the classical gravitational acceleration. So, the aim of this work is to show that when we consider that the total number of bits no longer satisfies the equipartition theorem then, at first, we can obtain two important results which are both the quantum correction of the gravitational acceleration and a MOND theory \cite{MOND1,MOND2,MOND3}.

In \cite{Pad1} the author brought the idea that we can deal mathematically with the concept that the spacetime can be considered an emergent structure.  Padmanabhan analyzed the idea that the Universe does not obey the holographic principle (asymptotic de Sitter Universe) and the consequence is the expansion of the Hubble volume as a function of the difference between the number of degrees of freedom of the holographic surface and the bulk, which follow the equipartition of energy law.

In this work we have discussed the emergence of spacetime through some aspects that we believe were not investigated so far.  One of these new aspects is the nonextensive one, where we have used the $q$-parameter as the underlying key to discuss the expansion issue.   This analysis is coupled to the solution of the main equation given in \cite{Pad1} considering matter and dark energy (DE) degrees of freedom.   After that we will discuss the degrees of freedom difference being represented by a general function and we have used, as an example, the one given by the Gauss-Bonnet (GB) gravity theory, in a $n$-dimensional generalization of our ideas.   We will obtain a new relation between the $q$-parameter and the GB coefficient, which brings interesting physical considerations.  The GB coefficient brings different physical features depending on its positive or negative value.  

In other words, we will obtain a constant parameter ($\alpha$) that is a measure of how much a difference between classical and quantum results concerning the number of bits is relevant.   We will see that this difference is not conserved at the quantum level.   In the second part of this work, the $\alpha$-parameter is a temperature function and after that a MOND's analysis will be provided.

This paper is organized in a way such that in section 2 we have depicted very briefly some Verlinde's entropic concepts.  In section 3 we have computed the $\alpha=N_{Sur}-N_E$ to show that $\alpha=0$ is not conserved at the quantum level.   In section 4 we have analyzed the emergence of cosmic space.  In section 5 we have introduced the nonextensive Tsallis concepts.   In section 6  we have carried out a $n$-dimensional generalization of the emergence of spacetime described here.
In section 7 the GB model in $n$-dimensions ($n \geq 4$) were discussed and in section 8 the MOND's modified gravity were investigated.
The conclusions and perspectives were written in section 9.

\section{Some Entropic gravity ideas}

Before we begin to describe our proposal, let us review, in a short way, Verlinde's formalism. The model considers a spherical surface as being the holographic screen, with a particle of mass $M$ located in its center. The holographic screen can be imagined as a storage device for information. The number of bits, which is the smallest unit of information in the holographic screen, is assumed to be proportional to the  holographic screen
area $A$
\begin{eqnarray}
\label{bits}
N = \frac{A }{l_P^2},
\end{eqnarray}

\ni where $ A = 4 \pi r^2 $ and $l_P = \sqrt{G\hbar/c^3}$ is the Planck length and $l_P^2$ is the Planck area.  We can see clearly that the connection between $l_P$ and $\hbar$ characterizes $l_P$ as a quantum parameter.  It means (at least) that its introduction can be considered a semi-classical feature.
In Verlinde's framework one can suppose that the bits total energy on the screen is given by the equipartition law of energy

\begin{eqnarray}
\label{eq}
E = \frac{1}{2}\,N k_B T.
\end{eqnarray}
It is important to notice that the usual equipartition theorem in Eq. (\ref{eq}), can be derived from the usual Boltzmann-Gibbs (GB) thermostatistics. 
Let us consider that the energy of the particle inside the holographic screen is equally divided by all bits in such a manner that we can have the expression

\begin{eqnarray}
\label{meq}
M c^2 = \frac{1}{2}\,N k_B T.
\end{eqnarray}
With Eq. (\ref{bits}) and using the Unruh temperature equation  \cite{unruh} given by

\begin{eqnarray}
\label{un}
k_B T = \frac{1}{2\pi}\, \frac{\hbar a}{c},
\end{eqnarray}
we are  able to obtain the  (absolute) gravitational acceleration equation

\begin{eqnarray}
\label{acc}
a &=&  \frac{l_P^2 c^3}{\hbar} \, \frac{ M}{r^2}\nonumber\\ 
&=& G \, \frac{ M}{r^2}\,\,.
\end{eqnarray}
From Eq. (\ref{acc}) we can see that Newton's constant $G$ is just written in terms of the fundamental constants, $G=l_P^2 c^3/\hbar$.

\section{Quantum gravitational correction through Verlinde's concepts}

It is well known that the infrared behavior of quantum gravity is more interesting than the ultraviolet one.  However, the low energy propagation of massless particles leads us to quantum corrections for long distances.  In this way, the effective action may be expanded in a Taylor momentum expansion.  In the current literature, the standard calculation procedure is through Feynman rules.  In a different way, in
this section we will use the holographic and equipartition law ideas to compute the quantum correction to the Newtonian potential.

We begin our formalism  by rewriting expression  (\ref{eq}) in a similar way as the one from Padmanabhan \cite{Pad1}, except by the fact that we consider just the holographic screen, as
\begin{eqnarray}
\label{class}
\frac{A}{l_p^2}=\frac{E}{(1/2) k_BT}.
\end{eqnarray}

\ni Then, our proposal is to assume that there is a numerical difference between the total number of bits in the holographic screen ($N_{Sur}$) and the number of bits that exist in the equipartition law ($N_E$)
\begin{eqnarray}
\label{dif}
\frac{A}{l_p^2}-\frac{E}{(1/2)k_BT}=\alpha,
\end{eqnarray}
or
\begin{eqnarray}
\label{dif2}
N_{Sur}-N_E=\alpha,
\end{eqnarray}
where 

\begin{eqnarray}
\label{dif3}
N_{Sur}\equiv\frac{A}{l_p^2},
\end{eqnarray}
and

\begin{eqnarray}
\label{dif4}
N_E\equiv\frac{E}{(1/2)k_BT}.
\end{eqnarray}

\ni Hence, as we have said before, we can understand that the $\alpha$-parameter measures the difference between the quantum $N_{Sur}$ and the classical $N_E$ objects.   However, since $N_{Sur}$ and $N_E$ are two objects that provide one quantity, namely, the number of bits, naively one can say that $\alpha \rightarrow 0$.   But in this section we will make two assumptions.   The first one is that $\alpha << 1$ but not zero.  And the second one, for the time being, is that $N_{Sur} > N_E$, namely, $\alpha > 0$.  We believe that both assumptions are not very different from reality.  Since they are simple assumptions, any result that contradicts any one of them will be easily noticed.   Here, in the future, when we will consider other backgrounds, we will make other assumptions about $\alpha$.  In advance, we will connect it to the Tsallis parameter and to the GB coefficient, where cosmological considerations will be discussed.

Initially if we consider that the $\alpha$  parameter is a constant then Eq. (\ref{dif}) can be written as

\begin{eqnarray}
\label{conta1}
k_BT&=&\frac{2E\, l_p^2}{A-\alpha l_p^2}\nonumber\\\nonumber\\
&=&\frac{2E l_p^2}{A} \(1- \frac{\alpha l_p^2}{A}\)^{-1}.
\end{eqnarray}

\ni Performing a binomial expansion in (\ref{conta1}) we obtain that
\begin{eqnarray}
\label{conta12}
k_BT=\frac{2E l_p^2}{A} \(1+ \frac{\alpha l_p^2}{A}+...\),
\end{eqnarray}
where we have assumed that $ \frac{\alpha l_p^2}{A}<< 1$. Using the Unruh temperature formula, 
Eq. (\ref{un}), and $E=M c^2$ in  (\ref{conta12}), we can obtain a modified gravitational acceleration 

\begin{eqnarray}
\label{conta2}
a=\frac{G M}{r^2} \(1+ \frac{\alpha}{4\pi} \frac{l_p^2}{r^2}+...\).
\end{eqnarray}

\ni The second term in Eq. (\ref{conta2}) is the first order non-relativistic quantum correction to the gravitational acceleration.
There are several papers that show different results for the coefficient $\alpha$ when general relativity is treated as an effective field theory. Among them we can mention the papers of Donoghue \cite{Don}, Akhundov et al \cite{Akh} and Bjerrum-Bohr et al \cite{BDH}.
As a simple potential cannot be considered an ideal relativistic concept, the general corrections would have the following expression

\begin{eqnarray}
\label{conta5}
V(r)=\frac{G m M}{r} \( 1+\beta \frac{l_p^2}{r^2}+... \),
\end{eqnarray}
where the $\beta$-parameter would rely on the exact definition of the potential [8].  It would be computed in the post-Newtonian expansion. The factor $l_p^2 /r^2$ is dimensionless and gives an expression parameter for the long distance quantum effects. The values of $\beta$ can be summarized in table I\footnote{Extracted from S. Faller, Theorieseminar Universit\"{a}t K\"{o}ln-02.02.2009.}

\begin{table}[h]
\center
\caption{}
\begin{tabular}{lc}
\hline
Work & $\beta$ \\
\hline
Donoghue  & $\;-\frac{127}{30\pi^2}(\approx -0.43)$\\
Akhundov et al.  & $\;-\frac{107}{30\pi^2}(\approx -0.36)$ \\
Bjerrum-Bohr et al.  &\;\;\; $\frac{41}{10\pi}(\approx 1.31) $ \\
\hline
\end{tabular}
\end{table}

\vskip 1 cm
\noindent We will use that the gravitational acceleration is connected to the potential through the equation 
$$ma=\left|\frac{dV}{dr}\right|,$$ then from (\ref{conta2}) and (\ref{conta5}) we can obtain a connection between $\beta$ and $\alpha$ given by

\begin{eqnarray}
\label{conta7}
\beta=\frac{\alpha}{12\pi}.
\end{eqnarray}

\ni Observing table I we can see that the first two negative results means that the number of bits that satisfy the equipartition theorem is greater than the total number of bits of the holographic screen. We consider that these results are not consistent with our proposal because we have assumed that $N_{Sur}\geq N_E$. Only the positive $\beta, (\beta=\frac{41}{10\pi})$ agrees with our model, i.e., $N_{Sur}\geq N_E$. We can imagine that $\alpha$ represents a small number of bits on the holographic screen. We have verified that if $\alpha=49$ then the value of $\beta$ obtained by Eq. (\ref{conta7}) approximately reproduces the value of $\beta$ given by Bjerrum-Bohr et al.  In fact, 49 bits is an extremely small value compared, as an example, with the number of bits on the holographic screen, Eq. (\ref{bits}), with radius 1m that is $\approx 10^{70}$. Therefore, $\alpha=49$ can be interpreted as a fluctuation in the equality of the bits number of the holographic screen and the bits number that satisfies the equipartition theorem. A  small deviation of equality (\ref{class}) leads to a quantum correction in the gravitational acceleration. It is important to mention here that in our proposal
we are not using any usual procedure of a quantum field theory or effective field theory for the gravitational interaction. Only the holographic principle and the equipartition theorem which are the basis of the Verlinde formalism were used.

In the next section, we will consider that the parameter $\alpha$ is the one which carries the information about the holographic principle, namely, if the holographic principle is obeyed or not.

\section{Emergence and expansion of the cosmic space}
\label{padmanabhan}

Let us deal again with the definition of the cosmic parameter $\alpha$ as in Eq. (\ref{dif2}), but in terms of the well known definition of the difference 
$\Delta N$, let us write that

\bee
\label{aaa}
\alpha\,=\,\Delta N\,=\,N_{Sur} \,-\, N_E
\eee

\ni where $N_E$ is the $N_{bulk}$ defined in \cite{Pad1}.  In this way we can say that the (from now on) cosmic parameter $\alpha$ is a measure of the expansion of the Universe, or the emergence of space \cite{Pad1}.  We will see in a moment in this section, that we do not consider $\alpha$ as being necessarily a constant parameter.  Let us use the definition \cite{Pad1} of the variation of cosmic space as being

\bee
\label{bbb}
\delta S_{av}\,=\,N_{Sur}\,-\,\epsilon N_E
\eee

\ni where, from \cite{Pad1} we have that, in $n\,=\,3$ space dimensions, that 

\bee
\label{ccc}
\delta S_{av}\,=\,\frac{1}{L^2_P}\,\frac{dV}{dt}\,=\,N_{Sur}\,-\,\epsilon N_E \,\,,
\eee

\ni where $V$ is the Hubble volume in Planck's units and $t$  is the cosmic time also in Planck's units.  $V$ is the volume of cosmic space surrounded by the apparent horizon, which is considered as being the marginally trapped surface with vanishing expansion.  Eq. (\ref{ccc}) is the dynamical equation for any phase and it can also be written as 

$$\frac{dV}{dt}\,=\,L^2_P\,(N_{Sur}\,+\,N_m\,-\,N_{DE})\,\,,$$

\ni where $N_m$ is the number of degrees of freedom of matter with $\rho + 3p\,>\,0\;(\epsilon=-1)$ and $N_{DE}$ is the number of degrees of freedom of DE with $\rho + 3p\,<\,0\;(\epsilon=+1)$.      Substituting $N_{Sur} $ given in Eq. (\ref{aaa}) into Eq. (\ref{bbb}) we have that

\bee
\label{ccc1}
\delta S_{av}\,=\,\alpha\,+\,(1-\epsilon)\,N_E\,\,,
\eee

\ni where $\epsilon=+1$ if $\rho+3p\,<\,0$ (accelerating phase), i.e., an asymptotic holographic equipartition, and $\epsilon=-1$ if $\rho+3p\,>\,0$.   So, $\delta S_{av}=\alpha$ if $\rho+3p\,<\,0$, we have an asymptotic holographic equipartition, and $\delta S_{av}=\alpha\,+\,2N_E$ if $\rho+3p\,>\,0$ for acceleration.   For $\alpha=0$ we have a pure de Sitter Universe and the holographic principle is obeyed ($N_{Sur}=N_E$) and $|E|\,=\,\frac 12\,N_{Sur}\,k_B\,T$ (holographic equipartition).   However, our real Universe is not pure de Sitter but it is an asymptotic de Sitter Universe, $\alpha \ne 0$, as we said above.
Using Eq. (\ref{ccc}) we can write that

\baa
\label{ccc2}
&&\frac{1}{L^2_P}\,\frac{dV}{dt}\,=\,\alpha\quad \nonumber \\
&&\mbox{} \\
&\Longrightarrow& \quad V\,=\,\alpha\,L^2_P \,t\,+\,V_0 \quad \mbox{if}\qquad\rho\,+\,3p\,<\,0\,\, \no
\eaa

\ni which, using Eq. (\ref{aaa}), it is equivalent to the standard Friedmann equation and $V_0$ means the current Hubble volume.   For $\alpha > 0$ in (\ref{ccc2}) we have the expansion of the Universe scenario since it grows with the cosmic time, i.e., we have DE degrees of freedom scenario, as expected.  For $\rho+3p\,>\,0$, we have that

\bee
\label{ddd}
\frac{1}{L^2_P}\frac{dV}{dt}\,=\,\alpha\,+\,2 N_E\,\,,
\eee

\ni where 
\bee
\label{ddd1}
N_{bulk}=N_E = \frac{2E}{k_B T}=\frac{2|\rho + 3p | V}{k_B T}\,\,,
\eee

\ni where $T$ is the horizon temperature $T=H/2\pi$ and $|E|\,=\,|\rho+3p|\,V$ is the Komar energy inside the Hubble volume $V=4\pi / (3H^3 )$.   The solution of Eq. (\ref{ddd}) is given by

\baa
\label{eee}
V\,=\,\frac{1}{\lambda} \Big[ (\alpha\,+\,\lambda V_0 )\,e^{\alpha L^2_P(t-t_0 )}\,-\,\alpha \Big] \,\,,
\eaa

\ni where $\rho\,+\,3p\,>\,0$ and $\lambda\,=\,\frac{4(\rho+3p)}{k_B T}$.  And Eq. (\ref{eee}) provides an expression for the expansion of the volume of spacetime as a function of the difference between the degrees of freedom since 
$\alpha\,=\,\Delta\,N$, besides time, of course.   Just below we will see a generalization of $\alpha$.
If $\alpha=-\lambda\,V_0$, in Eq. (\ref{eee}), it is easy to see that $V=\,-\,\alpha/\lambda =V_0$.   Hence, we have that 

\bee
\label{eee1}
N_{Sur}\,-\,N_E\,=\,\alpha\,=\,-\,\lambda\,V_0
\eee

\ni and since $\lambda>0$ (pure de Sitter Universe), $\rho=-p\:\Longrightarrow\: \lambda\,=\,8p/k_B T$, and $V_0 > 0$ we have that $\alpha < 0$ and $N_E\,>\,N_{Sur}$ and the holographic principle is not obeyed.   The number of degrees of freedom of the bulk is greater than the ones on the spherical surface of Hubble radius $H^{-1}$.   So, in this case of a pure de Sitter Universe ($\rho=-p$) we have that $\lambda=8p/(k_B T)\,>\,0$.

Since Verlinde stated that ``gravity can be identified with an entropic force caused by changes in the information associated with the position of material bodies," we will generalize and write Eq. (\ref{ccc2}) in terms of a function of $\alpha$, namely,

\bee
\label{iii}
\frac{1}{L_P^2}\,\frac{dV}{dt}\,=\,f(\alpha)\qquad \quad\mbox{if}\:\:\:\qquad \rho+3p < 0
\eee

\ni and consequently

\bee
\label{jjj}
V(t,\alpha)\,=\,f(\alpha)\,L_P^2\,t\,+\,V_0
\eee

\ni and, in the same way, from Eq. (\ref{ddd}) 

\bee
\label{kkk}
\frac{1}{L_P^2}\,\frac{dV}{dt}\,=\,f(\alpha)+2N_E \qquad\mbox{if}\qquad\rho+3p > 0 \,\,,
\eee

\ni which solution is

\bee \label{kkk1}
V(t,\alpha)\,=\,V_0 e^{f(\alpha) L_P^2 (t-t_0 )}\,-\,\frac{f(\alpha)}{\lambda} \Big[ 1\,-\,e^{f(\alpha) L_P^2 (t-t_0 )} \Big]
\eee

\ni where $\lambda$ is the same parameter as before.  We can see easily that the expansion of the Hubble volume can be written as a direct function of the cosmic parameter $\alpha$.  However, it is important to notice that this last expression is for a three space dimensional theory ($n=3$), which turns it not possible to consider the GB theory into this last solution.  As we said above, it would be necessary to generalize this last expression to $n\,>\,3$ spacetime dimensions, which will be done in a moment.




\section{Tsallis nonextensive considerations}
\label{tsallis}

Considering the results of \cite{Jan,aa}, using Tsallis thermostatistical approach, we have that the equipartition law can be given by

\bee
\label{mmm}
E\,=\,\frac{1}{5\,-\,3q}\,N\,k_B\,T \,\,,
\eee

\ni where $q$ measure the extensivity of the system and it is well known as the extensive parameter or Tsallis parameter.  With the following relation we can make the equivalence \cite{Jan,aa} computed in \cite{jiulin,jiulin2}

\bee
\label{mmm1}
T \, \longrightarrow \, \frac{2T}{5\,-\,3q}
\eee

\ni and we can write, from the $\lambda$ definition above, that

\bee
\label{fff}
\lambda\,=\,\frac{4(\rho\,+\,3p)}{k_B T} \: \longrightarrow \frac{2(5\,-\,3q)(\rho\,+\,3p)}{k_B T}
\eee

\ni and hence, for $\alpha\,=\,-\lambda\,V_0$, we have that

\bee
\label{ggg}
\alpha\,=\,-\,\frac{2(5-3q)(\rho+3p)}{k_B T}\,V_0\,\,.
\eee

\ni which is a new analysis in the nonextensive literature.  Namely, together with the new results obtained in the sections above, we have that, through this  equation (\ref{mmm1}) and consequently the last equation, we begin a new nonextensive analysis about some interesting and open questions such as the emergence of the cosmic spacetime and the GB gravity theory.

Hence, in Eq. (\ref{ggg}), if $q=5/3$ ($\alpha=0$) we have that the holographic principle is obeyed but from (\ref{eee}) we have that $\lambda=0$ is a singularity-type point concerning the Hubble volume.   So, we have that $\lambda \ne 0$ and there is not a specific value for $q$ where the holographic principle is obeyed.  We will see next that the $q=5/3$ value is a critical result but it is meaningless, physically speaking, of course.   However, inside an holographic analysis, we have a meaning for it, since it means that $\alpha=0$.

On the other hand, we can make a different analysis, namely, from Eq. (\ref{mmm}) we cannot have that $5-3q < 0$ and so, we must have that $q\,>5/3$.  Or, in other words, from (\ref{ggg}) it means that we cannot have that $\alpha>0\,\Longrightarrow\,N_{Sur}\,>\,N_{bulk}$.  Therefore, the only possible limit concerning $q$ is $q\,<\,5/3 \:(\alpha < 0)\:\Longrightarrow\,N_{Sur}\,< N_{bulk}$.   

Another way to explain this limitation in the $q$-parameter comes from \cite{Jan,aa} where it was explained that we have $G_{NE}=[(5-3q)/2]\,G_{grav.}$ (notice that, in this expression, when $q=1$ we have $G_{grav.}=G_{NE}$, i.e., BG scenario is recovered) and also that the value $q \geq 5/3$ makes no physical sense since we obtain $G_{grav.} \leq 0$.  So, $q < 5/3$ is an upper bound limit when we are dealing with the holographic principle.  Consequently, $\alpha < 0$ (matter degrees of freedom) is our upper bound limit concerning $\alpha$, since from (\ref{ccc2}) it means a contraction and not an expansion of the Universe and we have also that $(\rho+3p) > 0$ ($N_{Sur} < N_{bulk}$) represents matter degrees of freedom, as we have said before.  Therefore, since we can use the cosmic parameter $\alpha$ as a substitute for the relation $\rho+3p$, we can see directly $\alpha$ as the condition to have DE or matter degrees of freedom.   One can ask if we can do the same thing about the $q$-parameter, but it is limited in $q < 5/3$.

From (\ref{ggg}) we can write that

\bee
\label{hhh}
q\,=\,\frac 53 \,\bigglb[ 1\,+\,\frac{k_B T}{10 V_0 (\rho+3p)} \,\alpha \biggrb]
\eee

\ni and when the holographic principle is obeyed, $\alpha=0$, we have that $q=5/3$.  Hence, we have that $\alpha < 0$ and $q < 5/3$ means a matter scenario.
From Eq. (\ref{hhh}), to recover the BG background ($q=1$) we have that 

\bee \label{hhh1}
\alpha\,=\,-\,\frac{4(\rho+3p)V_0}{k_B T}\,\,,
\eee

\ni and the holographic principle cannot be obeyed, since all the terms in this last equation are different from zero.  To obey the holographic principle we have to consider a $T \longrightarrow \infty$ situation in (\ref{hhh1}).   The analysis is analogous to the one for Eq. (\ref{ggg}) for $q=1$, of course.

Having said that, we can see from (\ref{hhh1}) that the BG scenario is equivalent to the DE one, except when the holographic principle is obeyed at very high temperatures.  In this way, a thermodynamical analysis can be accomplished in this DE, holographic principle and BG structure, but it will not be considered here.

Back to the ideas given in \cite{Pad1}, from Eqs. (\ref{bbb}) and (\ref{ddd1}) we can write that

\bee
\label{az1}
N_{bulk}\,=\,-\,\epsilon\frac{2(\rho+3p)V}{k_B T} \,\,,
\eee

\ni where $\epsilon = -1$ means matter degrees of freedom and $\epsilon = 1$, the DE ones.   From Eq. (\ref{ddd1}), we have that

\bee
\label{az2}
N_{bulk}\,=\,-\,\frac{E}{\frac 12 k_B T}\,=\,\frac{2(\rho+3p)V}{k_B T}
\eee

\ni and if $\rho+3p < 0$, from (\ref{az1}) we have DE number of degrees of freedom.   So, let us use the EoS $p=\omega \rho$ in (\ref{az2}) to obtain that

\bee
\label{az3}
N_{bulk}\,=\,-\,\frac{2 (1+3\omega)V \rho}{k_B T}
\eee

\ni and, it is well known that, when $\omega\, <\, - 1 / 3$ we have DE degrees of freedom and if $\omega > 1 / 3$, matter degrees of freedom.   If the holographic principle is obeyed, we have that, $N_{Sur}=N_{bulk}$, where $N_{Sur}=A / L^2_P$, and $A$ is the area of the holographic surface of radius $H^{-1}$, i.e.,

\bee
\label{az4}
A\,=\,4\pi\, H^{-2} \quad \Rightarrow \quad N_{Sur}\,=\,\frac{4\pi}{L^2_P H^2}\,=\,\frac{2 (1+3\omega)V\rho}{k_B T} \,\,,
\eee

\ni which means DE degrees of freedom and that, using $T=H/(2\pi)$,

\bee
\label{az5}
H^2\,=\,\frac{4\pi(1+3\omega) \rho L^2_P}{k_B T}
\eee

\ni and substituting Eq. (\ref{mmm1}) into (\ref{az3}) we have that

\bee
\label{az6}
N_{bulk}\,=\,-\frac{ (1+3\omega)(5-3q)V\rho}{k_B T} \,\,,
\eee

\ni which shows us that for $\omega < - 1/3$ and from (\ref{hhh1}) that $\alpha > 0$ we will have DE degrees of freedom, since $q < 5/3$.
And using Eq. (\ref{mmm1}) into (\ref{az5}) we can write that

\bee
\label{az7}
H^2\,=\,\frac{2\pi (1+3\omega)(5-3q)\rho L^2_P}{k_B T}\,\,
\eee

\ni and, curiously, although in (\ref{az6}) the value $q=5/3$ also makes no sense, in Eq. (\ref{az7}) it means that we have no expansion at all 
since $$H=\frac{\dot{a}}{a}=0 \Rightarrow a (t)=\,constant\,\,,$$

\ni which makes no sense either.

\section{n-dimensional generalization}

Let us generalize the $D=3$ space dimensional ideas of the last section to $(n+1)$-spacetimes ($n\,>\,3$).   In Eq. (\ref{ccc2}), the generalization to $n$-space dimensions, is equal to

\bee
\label{z1}
\frac{d V}{d t}\,=\,L^{n-1}_P\,f (\alpha,N_{Sur})
\eee

\ni where, in $n$-dimensions we can write that $N_{Sur} = \beta A_n / L^{n-1}_P$, $A_n = n \Omega_n / H^{n-1}$ and $\Omega_{n}$ is the volume of the $n$ sphere.   Besides, $\beta = (n-1)/[2(n-2)]$ and $N_{bulk}$ is the number of degrees of freedom in the spherical volume given by $V\,=\,\Omega_n / H^n$.   So,

\bee
\label{z2}
N_{bulk}\,=\,\frac{|E|}{\frac 12 k_B T }\,=\,\frac{2}{k_B T}\,\frac{|(n-2)\rho\,+\,np|}{n-2}\,V\,\,,
\eee

\ni where $[|(n-2)\rho\,+\,np|V] / (n-2)$ is the $n$-dimensional bulk Komar energy.   Notice that in any equation from (\ref{z1}) till the end of this section we can reproduce the last section results when we substitute $n=3$.   Hence, let us choose the $f (\alpha, N_{Sur})$ as being the standard 

\bee
\label{z3}
f (\alpha,N_{Sur})\,=\, \frac{\alpha}{\beta}
\eee

\ni and from Eqs. (\ref{z1}) and (\ref{z3}) we can write that

\bee
\label{z34}
\frac{d V}{d t}\,=\,L^{n-1}_P \frac{\alpha}{\beta}
\eee

\ni where the $n$-dimensional $\alpha$ is $N_{bulk}-N_{Sur}$, so

\baa
\label{z35}
\alpha&=& \frac{\beta n \Omega_n}{L^{n-1}_P\,H^{n-1}}\,-\,\frac{2}{k_B T}\,\frac{|(n-2)\rho\,+\,np|}{n-2}\,V \nonumber \\
&=&\kappa_n\,-\,\bar{\kappa}_n\,V \nonumber \\
&=&\bar{\kappa}_n\,\Big(\frac{\kappa_n}{\bar{\kappa}_n}\,-\,V\Big) \,\,,
\eaa

\ni where 

\bee \label{z351}
\kappa_n=\frac{\beta n \Omega_n}{L^{n-1}_P H^{n-1}} 
\eee

\ni and

\bee \label{z352}
\bar{\kappa}_n=\frac{2}{k_B T}\frac{|(n-2)\rho+np|}{n-2}\,\,.
\eee

\ni Substituting Eq. (\ref{z35}) into (\ref{z34}) we have that

\bee
\label{z36}
\frac{d V}{d t}\,=\,\frac{L_P^{n-1}}{\beta}\,\bar{\kappa}_n\,\bigglb(\frac{\kappa_n}{\bar{\kappa}_n}\,-\,V\biggrb)
\eee

\ni and the solution of (\ref{z36}) is

\bee
\label{z4}
V\,=\,\bigglb(V_0\,-\,\frac{\kappa_n}{\bar{\kappa}_n}\biggrb)\, exp\bigglb [\frac{L^{n-1}_P\,\bar{\kappa}_n}{\beta}\, (t\,-\,t_0)\biggrb]\,+\,\frac{\kappa_n}{\bar{\kappa}_n}
\eee

Using again the transformation (\ref{mmm1})

\bee
T\,\longrightarrow\, \frac{2 T}{5-3q} \no
\eee

\ni in $\bar{\kappa}_n$, we can write that $\bar{\kappa}_n\,\longrightarrow\, (5-3q)\bar{\kappa}_n /2$ and substitute this expression in (\ref{z4}), the solution of (\ref{z36}) turns out to be

\baa
\label{z6}
V&=&\bigglb( V_0-\frac{2\kappa_n}{(5-3q)\bar{\kappa}_n}\biggrb)exp\bigglb[\frac{L_P^{n-1} (5-3q)\,\bar{\kappa}_n}{2\beta}\,(t\,-\,t_0)\biggrb] \nonumber \\
&\,+\,&\frac{2\kappa_n}{(5-3q)\bar{\kappa}_n}
\eaa

\ni and for $t=t_0$ we have that $V=V_0$, which is clearly a dimensional independent result, as expected.  Substituting in (\ref{z352})  the transformation for $T$ above, we can write explicitly that

\bee \label{z61}
\bar{\kappa}_n=\frac{5-3q}{k_B T}\frac{|(n-2)\rho+np|}{n-2}\,\,
\eee

\ni and, when $T \:\longrightarrow \: 0$ in this last equation, we have that $\bar{\kappa}_n \: \longrightarrow \: \infty$ which means, from (\ref{z35}), that $\alpha < 0$ and we have matter degrees of freedom.

In (\ref{z6}) let us analyze the initial condition

\bee \label{z62}
V_0\,=\,\frac{2\kappa_n}{(5-3q)\bar{\kappa}_n} \, \,,
\eee

\ni which means that $V=V_0$, namely, no expansion, hence matter degrees of freedom ($\alpha\, <\,0$).  

To calculate $q$ from this expression (\ref{z62}) we can write that

\bee
\label{z7}
q\,=\,\frac 53 \,-\,\frac{{2\kappa}_n}{3\bar{\kappa}_n\,V_0}
\eee

\ni where, from (\ref{z35}), i.e., using the definitions of $\kappa_n$ and $\bar{\kappa}_n$, it is easy to see from (\ref{z7}) that when $T \longrightarrow 0$ ($\bar{\kappa}_n \: \longrightarrow \: 0$) we have that $q=5/3$, the critical value, $\alpha = 0$ and the holographic principle is obeyed.  However, using this $\bar{\kappa}_n \:\longrightarrow \: \infty$ limit in Eq. (\ref{z61}) we have that $V_0 \; \longrightarrow \; 0$, which makes no sense.   So, we conclude that $T \longrightarrow 0$ is equivalent to $q \longrightarrow 5/3$, which is a singularity point.


\section{The  nonextensive analysis of the Gauss-Bonnet gravity}

Considering the generalization given in (\ref{iii}) and (\ref{kkk}) let us see a practical example of $f(\alpha)$ concerning a more complicated relation about the number of degrees of freedom on the surface $N_{Sur}$ and the number of degrees of freedom in the bulk $N_{bulk}$.   Let us use the function used in \cite{ylw}

\bee \label{z8}
f(\alpha , N_{Sur})\,=\,\frac{\frac{\alpha}{\beta}+\bar{\alpha} K_n B_n}{1+2\bar{\alpha}K_n C_n} \,\,,
\eee

\ni where $\bar{\alpha}$ is the well-known GB coupling coefficient of (length)$^2$ dimension,

\bee \label{z81} 
K_n\,=\,\bigglb(\frac{n\Omega_n}{L^{n-1}_P} \biggrb)^{\frac{2}{n-1}}
\eee

\ni and $\Omega_n$ is the volume of the $n$-sphere of unit radius.  We have also that

\bee \label{z82}
B_n\,=\,\bigglb(\frac{N_{Sur}}{\beta}\biggrb)^{1+\frac{2}{1-n}} \:,\quad C_n\,=\,\bigglb(\frac{N_{Sur}}{\beta}\biggrb)^{\frac{2}{1-n}}
\eee

\ni and in $n$-dimensions, again, we have

\bee \label{z83}
N_{Sur}\,=\,\frac{\beta A_n}{L^{n-1}_P} \:, \quad \beta = \frac{n-1}{2(n-2)} \:, \quad A_n = \frac{n\Omega_n}{H^{n-1}}\,\,.
\eee

\ni In the case of GB gravity, $n \geq 4$, otherwise, it reduces to a topological surface term and it has no dynamics.   
Considering the $f(\alpha)$ function, from Eq. (\ref{hhh}) we have for Tsallis parameter that

\bee
\label{nnn}
q\,=\,\frac 53 \,\bigglb[ 1\,+\,\frac{ k_B T}{10 V_0 (\rho+3p)} \,f(\alpha) \biggrb]\,\,.
\eee

\ni Hence, substituting (\ref{z8}) into (\ref{nnn}) we have that

\bee
\label{z9}
q(\alpha\,,\,\tilde{\alpha} )\,=\,\frac 53 \,\bigglb[ 1\,+\,\frac{ k_B T}{10 V_0 (\rho+3p)}
\frac{\alpha \,+\,\bar{\alpha}\beta K_n B_n}{\beta(1\,+\,2\bar{\alpha} K_n C_n )} \biggrb] \,\,
\eee

\ni and we can see a direct dependence of the $q$-parameter on the horizon temperature and the GB-parameter $\bar{\alpha}$.   We believe that this relation is a motivation to investigate the implications of the nonextensive concepts in GB gravity, which is an ongoing research.
And after some algebra, considering the BG scenario $(q=1)$ in (\ref{z9}), we have that

\bee \label{z10}
\alpha\,=\,-\,\bigglb[ \frac{4\beta}{k_B T}(\rho+3p) V_0 (1+2\bar{\alpha} K_n C_n )\,+\,\bar{\alpha}\beta K_n B_n \biggrb]
\eee

\ni and as we have said before, the bound for $q\,\,(q < 5/3)$ means the $\alpha < 0$ bound for $\alpha$.   In this way ($\alpha\, <\,0$), in (\ref{z10}) we have that

\baa \label{z11}
&&\frac{4\beta}{k_B T}(\rho+3p) V_0 (1+2\bar{\alpha} K_n C_n )\,+\,\bar{\alpha}\beta K_n B_n\,>\,0 \no \\
\mbox{} \no \\
&\Longrightarrow& \bar{\alpha}\, > \: 
-\,\Big(\frac{N_{Sur}}{\beta}\Big)^{\frac{2}{n-1}}\,\frac{1}{2K_n \Big[1+\frac{N_{Sur} k_B T}{8\beta V_0 (\rho+3p)} \Big]} \, \, , 
\eaa

\ni which means that we have an interval where the values for $\bar{\alpha}$ are negative.   It can be shown \cite{dehghani} that in GB gravity, negative values of the GB coefficient (without a cosmological constant), the acceleration of the expanding Universe can be explained, details in \cite{dehghani}.   So, from (\ref{z11}), we can see that, since the numerator is positive, the negative behavior of $\bar{\alpha}$ can be kept if the denominator is also positive, namely,

\baa \label{z111}
1\,+\,\frac{N_{Sur} k_B T}{8\beta V_0 (\rho+3p)}\,>\,0 \no \\
\mbox{} \no \\
\Longrightarrow \rho+3p \,>\, -\,\frac{N_{Sur} k_B T}{8\beta V_0}
\eaa

\ni and consequently, for the interval

\bee \label{z112}
-\,\frac{N_{Sur} k_B T}{8\beta V_0}\,< \rho+3p\,<\,0
\eee

\ni we have the confirmation that $\bar{\alpha} < 0$ means acceleration since condition (\ref{z112}) means DE degrees of freedom.   However, from (\ref{z11}) if 

\bee \label{z12}
1\,+\,\frac{N_{Sur}k_B T}{8\beta V_0 (\rho+3p)}\,<\,0\,\,,
\eee

\ni we have a positive GB coefficient.   From (\ref{z12}) we have that

\bee \label{z13}
\rho\,+\,3p\,<\,-\frac{N_{Sur}k_B T}{8\beta V_0}\,\,,
\eee

\ni which means also that we have DE degrees of freedom.   Notice that we are talking about a BG scenario.   For a positive GB coefficient it can be shown \cite{ess} that the GB term can play an anti-gravitation role in the cosmological context.  This anti-gravitation role under Tsallis nonextensive concepts is out of the scope of this paper, but it is another ongoing research.


\section{MOND's considerations}

The MOND theory successfully explains the majority of the rotation curves of the galaxies. MOND theory reproduces the well known Tully-Fisher relation \cite{TF} and it can be also an alternative to dark matter. 

This theory is a modification of Newton's second law in which the force can be described by

\begin{eqnarray}
\label{mond1}
F=m \, \mu \(\frac{a}{a_0}\) a,
\end{eqnarray}

\ni where $\mu(x)$ is a function which has the following properties: $\mu(x)\approx 1$ for $x>>1,\; \mu(x)\approx x$ for $x<<1$ and $a_0$ is a constant. There are different interpolating functions for $\mu(x)$ \cite{prof1,prof2}. However, it is believed that the main implications caused by MOND theory do not depend on the specific form of these functions. Therefore, for simplicity, it is usual to assume that the variation of $\mu(x)$ between the asymptotic limits occurs abruptly  at $x=1$ or $a=a_0$. 

Initially, let us consider \cite{Jan} that, below a critical temperature, the cooling of the holographic screen is not homogeneous. We choose that the fraction of bits with zero energy is given by the equation

\begin{eqnarray}
\label{mond2}
&\frac{N_0}{N}=1-\frac{T}{T_c},
\end{eqnarray}
where $N$ is the total number of bits given by the equation (\ref{bits}), $N_0$ is the number of bits with zero energy and $T_c$ is the critical temperature. For $T\geq T_c$ we have that $N_0=0$ and for $T\leq T_c$ the zero energy phenomenon for some bits starts to occur. Eq. (\ref{mond2}) is a standard relation of critical phenomena and second order phase transitions theory. From (\ref{mond2}) the number of bits with energy different from zero for a given temperature $T<T_c$ is

\begin{eqnarray}
\label{mond3}
N-N_0=N\frac{T}{T_c}.
\end{eqnarray}
Considering that the energy of the particle inside the holographic screen is equally distributed over all bits with nonzero energy and using relation (\ref{mond2}) in the equipartition law of energy, we obtain that

\begin{eqnarray}
\label{mond4}
Mc^2=\frac{1}{2} N \frac{T}{T_c} k_B T.
\end{eqnarray}
Then, combining (\ref{bits}), (\ref{un}) and (\ref{mond4}), we are able to derive, for $T<T_c$, the MOND theory for Newton's law of gravitation\footnote{In references\cite{Ep1} and \cite{Ep2} the authors derive MOND's theory by considering that the bits of the holographic screen are fermionic excitations.}

\begin{eqnarray}
\label{mond5}
a\(\frac{a}{a_0}\)=G\frac{M}{r^2},
\end{eqnarray}
where 

\begin{eqnarray}
\label{mond6}
a_0=\frac{2\pi c k_B T_c}{\hbar}.
\end{eqnarray}
Using $a_0\approx 10^{-10} m s^{-2}$ we obtain $T_c\approx 10^{-31} K$.

In order to generalize Eq.(\ref{mond2}), we will consider that

\begin{eqnarray}
\frac{N_0}{N}=1-\mu(x),
\end{eqnarray}
where
\begin{eqnarray}
x\equiv \frac{a}{a_0}=\frac{T}{T_c}.
\end{eqnarray}



\noindent Here we will use two of the most used interpolating functions \cite{prof2} in MOND theory that are the  ``simple" interpolating function

\begin{eqnarray}
\label{simple}
\mu(x)=\frac{x}{1+x},
\end{eqnarray}
and the ``standard" interpolating function

\begin{eqnarray}
\label{standard}
\mu(x)=\frac{x}{\sqrt{1+x^2}}.
\end{eqnarray}



Considering that there is a difference between the number of bits in the holographic screen and the number of bits that exists in the equipartition law, we can choose a particular expression for the $\alpha$-parameter, Eq. (\ref{dif}), that leads to MOND theory. To do this we will rewrite Eq. (\ref{mond4}) in the form

\begin{eqnarray}
\label{mond7}
N_{Sur}-N_E=N_{Sur}\[1-\mu\(\frac{T}{T_c}\)\],
\end{eqnarray}
or
\begin{eqnarray}
\label{mond8}
\frac{N_{Sur}-N_E}{N_{Sur}}=\[1-\mu\(\frac{T}{T_c}\)\],
\end{eqnarray}
where $N_{Sur}$ and $N_E$ is defined in Eqs. (\ref{dif3}) and (\ref{dif4}) respectively. Then, from Eq. (\ref{mond7}), the $\alpha$-parameter is given by
 
\begin{eqnarray}
\label{alfa}
\alpha=N_{Sur}\[1-\mu\(\frac{T}{T_c}\)\],
\end{eqnarray}

\ni which shows that in the MOND theory the $\alpha$-parameter depends on the temperature.

In Figure 1 we have plotted $(N_{Sur}-N_E)/N_{Sur}$ as function of $T/T_c$, where we have used the two interpolating functions, Eqs. (\ref{simple}) and (\ref{standard}). We can mention that for $\frac{T}{T_c} << 1$ we have the MOND behavior dominance, and for 
$\frac{T}{T_c} >> 1$ we have the standard gravity behavior dominance. We can also mention that when varying $T/T_c$, the expression $(N_{Sur}-N_E)/N_{Sur}$ approaches to the standard gravity($(N_{Sur}-N_E)/N_{Sur}=0$) more rapidly if we use the standard interpolating function, Eq. (\ref{standard}).

\begin{figure}[ih]
\includegraphics[scale=.7,angle=0]{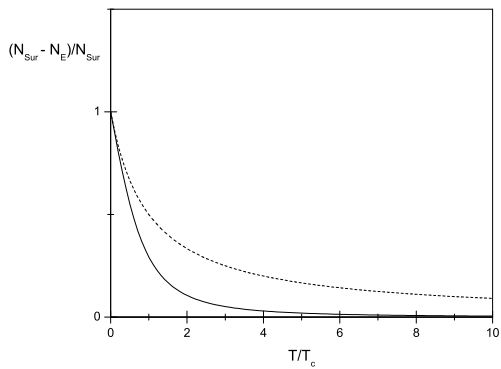}
\caption{ $ \frac{N_{Sur}-N_E}{N_{Sur}}$ as function of $T/T_c$. The Dash line represents the ``simple" interpolating function, Eq. (\ref{simple}), being used in Eq. (\ref{mond8}). The Solid line represents the ``standard"' interpolating function, Eq. (\ref{standard}), being used in Eq. (\ref{mond8}).}
\end{figure}

\section{conclusions}

Two recent formalisms that were originated in order to introduce alternative models concerning gravity theory, the so-called MOND and Verlinde's entropic gravity promoted a whole scientific research Universe. The first one had the objective to explain the galaxies' rotation curves and it was considered as an alternative to dark matter.  The second one was aimed to continue the ideas of Beckenstein and Hawking who explored the black holes thermodynamics through entropic concepts and the Unruh thermodynamical acceleration definition and in the end we can obtain Newton's gravitational law.

In this work, we have used both frameworks to obtain an $\alpha$-parameter which has worked as a measure for classical/quantum differences and to analyze the phase transition and critical phenomena.   What is new here about these results is that both are considered on the holographic screen, which is an important concept in the current information theory.

The first result concerning the $\alpha$-parameter has shown that the difference between the number of bits on the holographic screen ($N_{Sur}$) and the equipartition law ($N_E$) is not zero, as one would expect.   In other words, since we have shown also that $\alpha$ has quantum features, at the quantum level, $N_{Sur}\,-\,N_E$ is not conserved, which could be considered as a kind of anomaly.   The $\alpha$-parameter has also appeared as the perturbative correction of the Newton law and we have compared it with other values obtained in the literature.  We have seen that matter and DE cosmological backgrounds can be connected to this cosmic parameter $\alpha$.  After that, we have made a bridge between this cosmic parameter and the Tsallis nonextensive one.

Interesting results had appeared when we introduced these ideas in the GB gravity theory.  We have discussed the influence of Tsallis concepts and the holographic ones in order to obtain cosmological consequences through the analysis of the GB coefficient signal, namely, the fact that it can be  negative or positive, meaning different cosmological phenomena and its connection to the other results obtained here.

In other words, we have tried to change the point of view of the parameters used in the literature to determine if we have matter or DE scenarios.  The cosmic parameter could be the used one instead of the equation of state, for example.  



\section{Acknowledgments}

\ni E.M.C.A. thanks CNPq (Conselho Nacional de Desenvolvimento Cient\' ifico e Tecnol\'ogico), Brazilian scientific support federal agency, for partial financial support through Grants No. 301030/2012-0 and No. 442369/2014-0 and the hospitality of Theoretical Physics Department at Federal University of Rio de Janeiro (UFRJ), where part of this work was carried out. 




\end{document}